\documentclass[preprint,proceedings]{rmaa}

\newcommand{\D}{\discretionary{}{}{}}

\SetYear{2002}
\SetConfTitle{Winds, Bubbles and Explosions}

\title{Dust Formation Events in Colliding Winds: {\it an application to $\eta $} Car}

\author{D. Falceta-Gon\c calves\altaffilmark{1}, V. Jatenco-Pereira\altaffilmark{1} \&
 Z. Abraham\altaffilmark{1}}

\altaffiltext{1}{ Instituto de Astronomia, Geof\'\i sica e Ci\^encias Atmosf\'ericas - 
Universidade de S\~ao Paulo \\
 Rua do Mat\~ao, 1226 - 05508-090 - S\~ao Paulo - Brazil 
(diego\D{}@astro.\D{}iag.\D{}usp.\D{}br).}

\suppressfulladdresses

\listofauthors{D. Falceta-Gon\c calves}
\indexauthor{Falceta-Gon\c calves, D.}

\addkeyword{binary systems}
\addkeyword{dust formation}
\addkeyword{Stars: Mass loss}

\begin{document}
% Typeset article header
\maketitle 

\boldabstract{Recent IR observations indicate that many massive binary systems 
present dust formation episodes, in regions close to the stars, during the {\it periastron} 
passage. 
These systems are known to be high-energy sources, and it is believed that wind 
collisions are the origin of the emission. In this work we show that wind collisions 
not only increase the X-ray emission but also allow dust formation. As an 
application we study $\eta$  Car, which presents, near {\it periastron}, an increase in the X-ray 
emission followed by a sudden decrease that lasts for about 
a month. We  reproduce this 
feature calculating the optical depth due to dust formation along the orbital period.}
\smallskip

Individual massive stars emit, in general,  $\sim 10^{33}$ erg s$^{-1}$ in X-rays , but 
binary systems are known to emit $\sim 100$ times more in this band. This discrepancy 
indicates that a large amount of this energy comes from a non-stellar source. 
Usov (1992) developed an X-rays emission model for colliding winds in massive 
binary systems in which the  gas becomes denser and hotter ($\sim 10^{8}$ K) in the shocked region, 
increasing the free-free emission in the X-ray band. However, some systems, as $\eta$ Car,
present anomalous light curves, with sudden 
decreases in flux, which rises again after some period of low emission (Ishibashi {\it et al.} 1999). 
Many massive binary systems also present high IR 
emission ($e.g.$ Monnier, Tuthill \& Danchi 2001), which is associated with dust, 
formed close to the stars mainly during the {\it periastron} passage.

Shocks  between winds will be occurring during all the orbital period, but near 
{\it periastron} they will be stronger. 
The resulting hot and dense shocked gas will emit large amounts of energy, cooling in a short time 
scale ($\sim$ few hours) from $10^{8}$ K to just $10^{4}$ K. As the gas cools, it also 
becomes  denser generating an  optically thick screen  to the ionizing photons around the system.
The neutral region behind the screen becomes even cooler ($\sim 10^{3}$ K). At this temperature 
and density dust may form and grow fast increasing the high-energy absorption.
 
We applied the wind collision model for the  $\eta$ Car binary system, assuming stellar 
parameters for $\eta$ Car given by Corcoran {\it et al.} (2001) and a WR  as the companion star,
with stellar parameters given by $e.g.$ Lamers (2001).  At {\it periastron} the dense region
reaches approximately $10^{13}$ cm, the ionized part only $10^{9}$ cm. The temperaure of the
neutral region varies between $\sim 1800$ K, at the boundary with the 
ionized region to $\sim 100$ K at the external radius. Graphite dust grains grow  to an equilibrium 
size of  $0.1\; \mu{\rm m}$ in about 5 hours. The optical 
depth in $2 - 10$ keV band is given by:
%\begin{equation}
%\smallskip
$$\tau_{X} \simeq 10^{3}\times \left(\frac{\dot{M}(\rm {M}_{\odot}/\rm {yr})}{10^{-3}}
\right) \left(\frac{100}
{v (\rm {km/s})}\right) \left(\frac{1}{R(\rm {A.U.})}\right),$$
%\smallskip
%\end{equation}
\par\noindent  
allowing a great decrease in the observable X-ray flux. Assuming an eccentricity 
for the system of $e = 0.8$ and a mass loss rate of $\dot{M}=3\times 10^{-4}\;
\rm {M}_{\odot} \; \rm {yr}^{-1}$, we could reproduce the strong X-ray flux absorption
near {\it periastron} passage. The later increase in X-ray emission is  simply due to expansion
of the dust cloud. This model 
is also in agreement with recent observations of several massive binary systems, 
as WR137, WR134, WR125 and WR140 (Williams, Kidger \& van der Hucht 2001; 
Kwok, Volk \& Bideldman 1997; Monnier, Tuthill \& Danchi 2001).

%\smallskip
\acknowledgements
This work was partially supported by the Brazilian Agencies CNPq, FAPESP, and
FINEP.
%\smallskip

\end{document}